\documentclass[letterpaper,times]{sae}

\usepackage[svgnames]{xcolor}
\definecolor{SAEblue}{RGB}{1,160,233}
\raggedright

\usepackage[justification=justified, singlelinecheck=false]{caption}  
\DeclareCaptionFont{eightpt}{\fontsize{8pt}{11pt}\selectfont #1}
\usepackage{lineno}
\usepackage[utf8]{inputenc}

\usepackage{array}
\newcolumntype{L}[1]{>{\raggedright\let\newline\\\arraybackslash\hspace{0pt}}p{#1}}
\newcolumntype{C}[1]{>{\centering\let\newline\\\arraybackslash\hspace{0pt}}p{#1}}
\newcolumntype{R}[1]{>{\raggedleft\let\newline\\\arraybackslash\hspace{0pt}}p{#1}}

\usepackage{graphicx}
\usepackage{multirow}
\usepackage{amsmath}
\usepackage{todonotes}
\usepackage[utf8]{inputenc}
\usepackage{amsmath,amssymb,amsfonts}
\usepackage{textgreek}
\usepackage{graphicx}
\usepackage{color,bm}
\usepackage{wasysym}
\usepackage{tikz}
\usepackage{cite}
\usepackage{array,multirow}
\usepackage{float}
\usepackage{balance}
\usepackage{hyperref}
\hypersetup{hidelinks}
\usepackage{mathtools, nccmath}
\DeclareGraphicsExtensions{.tif}
\usepackage[linesnumbered,ruled,vlined]{algorithm2e}
\usepackage{amsmath} 
\usepackage{algorithmic,algorithm2e}

\usepackage{nameref}
\usepackage{booktabs}

\makeatletter
\def\@seccntformat#1{%
  \expandafter\csname c@#1\endcsname\c@section
  }
\makeatother

\usepackage{subfigure}
\usepackage{multimedia}

\PaperTitle{Motion Estimation of Connected and Automated Vehicles under Communication Delay and Packet Loss of V2X Communications}
\usepackage{calc} 

\makeatletter 
\renewcommand\@biblabel[1]{#1. } 
\makeatother

\AddAuthor{Ziran Wang, Kyungtae Han, and Prashant Tiwari}{Toyota Motor North America R\&D, InfoTech Labs, Mountain View, CA}
\PaperNumber{2021-01-0202}
\SAECopyright{2021}

\begin{document}
\maketitle
\section{Abstract}
The emergence of the connected and automated vehicle (CAV) technology enables numerous advanced applications in our transportation system, benefiting our daily travels in terms of safety, mobility, and sustainability. However, vehicular communication technologies such as Dedicated Short-Range Communications (DSRC) or Cellular-Based Vehicle-to-Everything (C-V2X) communications unavoidably introduce issues like communication delay and packet loss, which will downgrade the performances of any CAV applications. In this study, we propose a consensus-based motion estimation methodology to estimate the vehicle motion when the vehicular communication environment is not ideal. This methodology is developed based on the consensus-based feedforward/feedback motion control algorithm, estimating the position and speed of a CAV in the presence of communication delay and packet loss. The simulation study is conducted in a traffic scenario of unsignalized intersections, where CAVs coordinate with each other through V2X communications and cross intersections without any full stop. Game engine-based human-in-the-loop simulation results shows the proposed motion estimation methodology can cap the position estimation error to $\pm 0.5$ m during periodic packet loss and time-variant communication delay. 

\section{Introduction}
\subsection{Background}
During the past decades, the emergence of connected and automated vehicle (CAV) technology brings new possibilities to our transportation systems. Specifically, the level of connectivity within our vehicles has greatly increased, allowing these ``equipped'' vehicles to behave in a cooperative fashion. The cooperation is not only among vehicles themselves through vehicle-to-vehicle (V2V) communications, but also among vehicles and other transportation entities through vehicle-to-infrastructure (V2I) communications, vehicle-to-network/cloud (V2N/V2C) communications, vehicle-to-pedestrian (V2P) communications, and etc. In summary, all these communication technologies are categorized as vehicle-to-everything (V2X) communications.

In the system architecture of a CAV, the communication module plays an equally important role as other modules, such as the localization module, perception module, planning module, and control module. It facilitates real-time and reliable wireless V2X communications among CAVs and other entities, either through Dedicated Short-Range Communications (DSRC) or Cellular-Based Vehicle-to-Everything (C-V2X) communication technologies. The communication module of a CAV system is capable of providing additional information that cannot be readily detected by the perception module, and can generally provide information more quickly than through sensor detection and processing. These scenarios may include the following:

\begin{itemize}
    \item Information from other CAVs that are beyond perception ranges of the ego CAV, or that are occluded from view by intermediate vehicles or by horizontal/vertical road curvatures.
    \item Information from other CAVs that cannot be directly sensed by the perception module, such as vehicle status information (e.g., wheel speeds, fault status, performance capabilities, etc.).
    \item Immediate notification of speed changes or steering commands as soon as they have been issued to other CAV’s actuators, even before their motions have begun to change.
    \item Negotiations between CAVs regarding desired maneuvers (e.g., merging, lane changing), so that these can be done more safely and efficiently.
\end{itemize}

Given the strength of information sharing among CAVs through their communication modules, cooperative automated driving can be enabled in multiple traffic scenarios, such as Cooperative Adaptive Cruise Control (CACC) \cite{shladover2015cooperative, wang2018areview}, cooperative ramp merging \cite{riostorres2017asurvey, wang2019cooperative}, speed harmonization on highways \cite{ma2016freeway, malikopoulos2019optimal}, cooperative eco-driving at signalized intersections \cite{altan2017glidepath, wang2020cooperative}, and automated coordination at unsignalized intersections \cite{dresner2008amultiagent, guney2018scheduling}. In these applications of cooperative automated driving, CAVs are coordinated to maintain safe inter-vehicle distances and accomplish tasks together based on V2X communications.

However, in a wireless communication network, due to the uncertain reliability of wireless communication links, communication time delays and packet losses are always unavoidable in the information sharing process among CAVs. Along with such time-delayed and partially dropped measurements of vehicles' motions, the control module of a CAV system needs to be carefully designed to reduce the adverse impact of vulnerable feedback channels on the system performance.

\subsection{Literature Review}

To address this time delay issue of V2X communications, two typical methods for delayed system (Razumikhin-based method and Krasovskii-based method) were studied in the literature. For example, Gao et al. considered uniform and constant time delays with a Razumikhin-based method applied to synthesize an H-infinity controller \cite{gao2016robust}. Besides uniform and constant time delays, Petrillo et al. considered multiple time-varying delays and used adaptive feedback gains to compensate for the errors arising from outdated information \cite{petrillo2018adaptive}. Di Bernardo et al. designed a consensus-based CACC controller by using the concept of aggregate delay, and a Razumikhin-based method was applied to the stability analysis \cite{bernardo2015distributed}, which was further extended to the case of heterogeneous time delays in \cite{bernardo2016design}. To compare the Razumikhin-based and Krasovskii-based methods, Chehardoli et al. further designed a consensus-based controller and demonstrated the less conservatism of the latter method in the sense of a greater upper bound of time delays \cite{chehardoli2018third}.

As for packet losses studied in the literature, one feasible solution is to achieve smooth transition to remove the reliance on V2X communications. For example, Ploeg et al. designed an acceleration estimation algorithm using on-board sensors in the case of communication failures so as to achieve seamless transition from CACC to Adaptive Cruise Control (ACC) \cite{ploeg2015graceful}. Harfouch et al. designed an adaptive switched controller for the transition from CACC to ACC to address the network switching due to communication failures \cite{harfouch2017adaptive}.

Quite a few works have been done regarding motion estimation of vehicles with sensor-fusion method, such as Extended Kalman filter (EKF) \cite{wenzel2006dual}, unscented Kalman Filter (UKF) \cite{antonov2011unscented}, and particle filter \cite{bogdanski2018kalman}. These filters are the modified versions of Kalman filter, and are based on the first-order Markov chain, namely the state at current time step is conditioned only on the state at the previous time step. Although they are useful for real-time implementation, a crucial limitation is that sensor data prior to the previous time step cannot be directly used. According to that, moving horizon estimation methodologies have been proposed, which utilize all data acquired in previous time steps, and can handle irregular sampling rates and delayed data without any additional modifications \cite{zanon2013nonlinear, mori2020simultaneous}. However, all aforementioned works have been primarily focused on the motion estimation of the ego vehicle, where data acquired from various sensors are fused to provide the estimation result. To the best of the authors' knowledge, very few literature has dealt with motion estimation of a target connected vehicle without the help of additional sensors.

\subsection{Contributions of the Paper}

Compared to the existing literature that studied the communication delay and packet loss aspects of V2X communications, we make the following contributions in this paper:

\begin{itemize}
    \item Different from most existing literature, we consider both communication delay and packet loss aspects in our motion estimation algorithm to design a reliable and robust CAV system. 
    
    \item We develop a consensus-based motion estimator and controller that allow the cooperative automated driving system to keep functioning, without necessarily degrading it to a non-cooperative system that is much less advanced like \cite{ploeg2015graceful, harfouch2017adaptive}.
    
    \item We conduct a case study of automated coordination at unsignalized intersections, where CAVs cooperate with each other coming from different legs of an intersection and cross without any full stops.
    
    \item Instead of only conducting numerical simulations to prove the effectiveness of the system, we measure V2X communication parameters by real connected vehicles, and then develop a high-fidelity simulation platform based on Unity game engine.
\end{itemize}

In the rest of this paper, we firstly formulate the problem to solve in this paper, covering system assumptions, system preliminaries, and problem statement. Then, we propose the methodology of the paper, which is the consensus-based motion estimation methodology for CAVs under communication delay and packet loss of V2X communications. Next, the game engine-based simulation is conducted, which validates the effectiveness of the proposed methodology. Finally, we draw the conclusion of this paper, and point out some future steps that can further improve this study.

\section{Problem Formulation}

\subsection{System Assumptions}

As described earlier, CAVs rely on on-board perception sensors, such as camera, radar, and LIDAR, to measure neighboring vehicles’ states. With the introduction of V2X communications, CAVs can obtain the states of those beyond their direct measurement ranges and to obtain information that cannot be detected by remote sensors (such as the issuance of internal control commands). This helps enhance the sensing range of CAVs, but also brings about various communication issues to CAV systems. Fortunately, because the communicated information is supplementary to the information obtained from on-board perception sensors, it is possible for cooperative automated driving systems to be degraded to non-cooperative automated driving systems when communication issues occur. However, this system degradation unavoidably decreases the effectiveness of the original CAV systems, since information flows among different vehicles will disappear, and CAV systems will become automated vehicle (AV) systems.

In this paper, we develop a motion estimation methodology that still enables cooperative automated driving even when communication issues occur. Information flows among CAVs still exist ``virtually'', which are based on the estimated motions of CAVs, instead of the ground-truth motions measured by the motion sensors of CAVs. Whenever the V2X communication recovers from the undesired status (e.g., communication delay decreases to a normal range, or packet loss disappears), those information flows among CAVs will return to the ground-truth motions. With the proposed motion estimation methodology, information flows are working all the time in the CAV systems, avoiding any system degradation from cooperative automated driving to non-cooperative automated driving.

It shall be noted that, since this paper mainly focuses on designing the motion estimation methodology, some reasonable specifications and assumptions are made while modeling the system to enable the theoretical analysis:

\begin{itemize}
    \item Only the longitudinal motion of vehicles is considered in this paper, where the motion control algorithm and motion estimator are both designed for the longitudinal vehicle dynamics. 
    \item The focus of this paper is estimating the vehicle motion under observed/measured V2X communication issues. The reasons that cause such V2X communication issues are not studied in this paper.
    \item A specific V2X communication technology (either DSRC or C-V2X) is not required for our proposed method. The motion estimation methodology is supposed to handle issues caused by any V2X communication technologies.
    \item Although the perception module of a CAV (i.e., camera, radar, and/or LIDAR) can provide auxiliary information when the performance of V2X communications is impaired, we do not consider its measurements in our motion estimation methodology due to the scope of this paper.
\end{itemize}

\subsection{System Preliminaries}

First, given the longitudinal dynamics of a vehicle $i$ as the following equations:

\begin{equation}
\begin{aligned}
\dot{r}_{i}(t) &=v_{i}(t) \\
\dot{v}_{i}(t) &=a_{i}(t) \\
a_{i}(t) &=\frac{1}{m}\left[F_{n e t_{i}}(t)-R_{i} T_{b r_{i}}(t)-c_{v i} v_{i}(t)^{2}-c_{f i} v_{i}(t)-d_{m i}(t)\right]
\end{aligned}
\end{equation}

where ${r}_{i}(t)$, $v_{i}(t)$, and $a_{i}(t)$ denote the longitudinal position, longitudinal speed and longitudinal acceleration of vehicle $i$ at time $t$, respectively, $m_i$ denotes the mass of vehicle $i$, $F_{n e t_{i}}$ denotes the net engine force of vehicle $i$ at time $t$, which mainly depends on the vehicle speed and the throttle angle, $R_i$ denotes the effective gear ratio from the engine to the wheel of vehicle $i$, $T_{b r_{i}}(t)$ denotes the brake torque of vehicle $i$ at time $t$, $c_{vi}$ denotes the coefficient of aerodynamic drag of vehicle $i$, $c_{fi}$ denotes the coefficient of friction force of vehicle $i$, $d_{mi} (t)$ denotes the mechanical drag of vehicle $i$ at time $t$.

We can then derive the following equations from the principle of vehicle dynamics when the braking maneuver is deactivated, i.e., vehicle $i$ is accelerating by the net engine force:

\begin{equation}
F_{n e t_{i}}(t)=\ddot{x}_{i}(t) m_{i}+c_{v i} \dot{x}_{i}(t)^{2}+c_{p i} \dot{x}_{i}(t)+d_{m i}(t)
\end{equation}

and we have the following equation when the braking maneuver is activated (vehicle $i$ decelerates by the brake torque):

\begin{equation}
T_{b r_{i}}(t)=\frac{\dot{x}_{i}(t) m_{i}+c_{v i} \dot{x}_{i}(t)^{2}+c_{p i} \dot{x}_{i}(t)+d_{m i}(t)}{R_{i}}
\end{equation}

Note that the net engine force is a function of the vehicle speed and the throttle angle, which is typically based on the steady-state characteristics of engine and transmission systems. The associated mathematical derivation can be referred to \cite{xiao2011practical}.

Based on the existing literature \cite{milanes2014cooperative}, the motion control module of a CAV is based on a hierarchical strategy, where the high-level controller generates a target acceleration (the first two equations in (1)), while the low-level controller commands the vehicle actuators to track the target acceleration (the last equations in (1)). In this paper, we focus on the high-level vehicle controller, where we propose the motion estimator based on the dynamics consensus of CAVs. 

\subsection{Problem Statement}

\begin{figure}[!htb]
\centering
\includegraphics[width=0.45\textwidth]{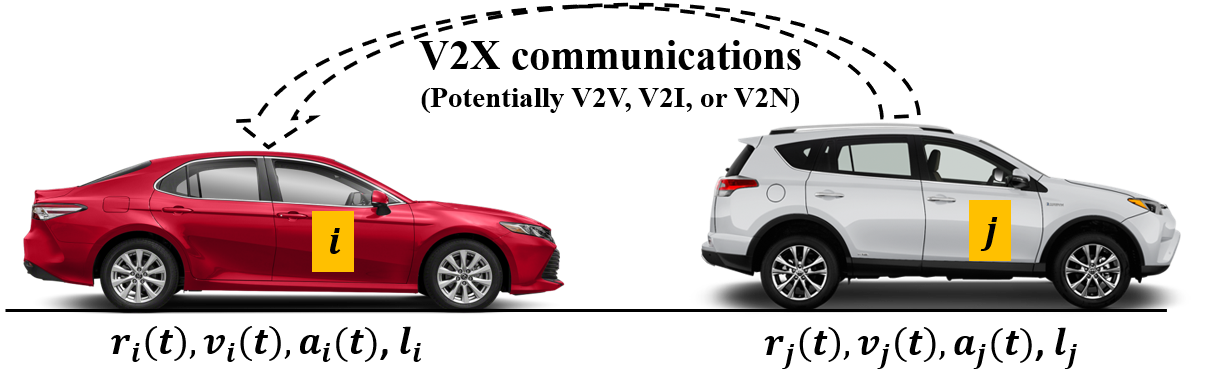}
\caption{The illustration of vehicle parameters in a V2X communication environment with an ego vehicle $i$ and its target vehicle $j$.}
\label{fig:dynamics}
\end{figure}

Given a two-CAV scenario, where the ego vehicle $i$ gets information from its target vehicle $j$ through V2X (e.g., V2V, V2I, or V2N) communications, the dynamics consensus can be generalized as a longitudinal control problem shown as Fig. \ref{fig:dynamics}.

In this figure, $r$, $v$, $a$, and $l$ denote the vehicle longitudinal position, longitudinal speed, longitudinal acceleration, and length, respectively. Simply speaking, the dynamics consensus of CAVs can be demonstrated as follows:

\begin{equation} \label{eq:converge}
\begin{array}{l}
r_{i}(t) \rightarrow r_{j}(t)-r_{\text {headway}} \\
v_{i}(t) \rightarrow v_{j}(t) \\
a_{i}(t) \rightarrow a_{j}(t)
\end{array}
\end{equation}

where $r_{\text{headway}}$ denotes the desired distance headway between these two vehicles. Although this dynamics consensus does not apply to all traffic scenarios (where some of them might require the ego vehicle to have a higher/lower speed/acceleration than the target vehicle), it can be applied to most of the cooperative longitudinal motion control applications of CAVs, such as the ones mentioned in the first section of this paper (e.g., CACC, speed harmonization, etc.).

However, the cooperative longitudinal motion control of CAVs heavily relies on the performance of V2X communications, which enable the information transmission between these two vehicles shown in Fig. \ref{fig:dynamics}. Time delay \cite{gao2016robust, petrillo2018adaptive} and packet loss \cite{ploeg2015graceful, harfouch2017adaptive} are two major issues that impair the performance of V2X communications in CAV applications. Different from most of the existing literature that tackle these two issues separately, we propose a motion estimation methodology in this paper to address them at the same time. 

\section{Methodology}

\subsection{Consensus-Based Motion Control}

In order to achieve the dynamics consensus of CAVs in equation (4), a double-integrator consensus-baesd longitudinal motion control algorithm can be proposed as

\begin{equation} \label{eq:consensus}
\begin{array}{l}
\dot{r}_{i}(t)=v_{i}(t) \\
\dot{v}_{i}(t)=-\alpha_{i j} k_{i j} \cdot\left[\left(r_{i}(t)-r_{j}(t)+l_{j}+v_{i}(t) \cdot t_{i j}^{g}(t)\right)+\gamma_{i}\right.\cdot \\
\left.\left(v_{i}(t)-v_{j}(t)\right)\right]
\end{array}
\end{equation}

where $\alpha_{i j}$ is the adjacency matrix of the directed graph (i.e., V2X communication topology between vehicle $i$ and $j$), $t_{i j}^{g}(t)$ is the time-variant desired time gap between two vehicles, which can be adjusted by many factors like road grade, vehicle mass, braking ability, etc. The term $[l_{j}+v_{i}(t) \cdot t_{i j}^{g}(t)]$ is another form of the term $r_{\text{headway}}$ in equation (\ref{eq:converge}). The control gains $k_{ij}$ and $\gamma_i$ in this algorithm can be either defined as constants, or further tuned by a feedforward control algorithm to guarantee the safety, efficiency, and comfort of this slot-following process. A lookup-table approach is adopted to dynamically calculate these control gains, based on the initial speeds of two vehicles, as well as their initial headway. In short, it can be summarized as

\begin{equation}
    \{k_{ij}, \gamma_i\} = f\big(v_i(0), v_j(0), r_i(0) - r_j(0)\big)
\end{equation}

\noindent where the details can be referred to our previous work \cite{wang2019lookup}. With this longitudinal motion control algorithm equation (\ref{eq:consensus}), vehicle $i$ in Fig. \ref{fig:dynamics} is able to converge its longitudinal speed $v_{i}(t)$ to vehicle $j$'s longitudinal speed $v_{j}(t)$, and converge its longitudinal position ${r}_{i}(t)$ to vehicle $j$'s longitudinal position ${r}_{j}(t)$ minus a desired headway between them.

However, the aforementioned algorithm does not consider the communication delay issue. It is without a doubt that, whatever information vehicle $i$ receives is not the exactly current information of vehicle $j$, due to the unavoidable transmission time $\tau_{ij}(t)$. Therefore, equation (\ref{eq:consensus}) can be further written into the following form while considering the time-variant communication delay:

\begin{equation} \label{eq:consensusdelay}
\begin{array}{l}
\dot{r}_{i}(t)=v_{i}(t) \\
\dot{v}_{i}(t)=-\alpha_{i j} k_{i j} \cdot\left[\left(r_{i}(t)-r_{j}(t-\tau_{ij}(t))+l_{j}+v_{i}(t) \cdot t_{i j}^{g}(t))\right)+\gamma_{i}\right.\cdot \\
\left.\left(v_{i}(t)-v_{j}(t-\tau_{ij}(t))\right)\right]
\end{array}
\end{equation}

\subsection{Motion Estimation for Communication Delay and Packet Loss}

As stated in the problem statement of this paper, time delay and packet loss are two major issues that impair the performance of V2X communications in CAV applications. In this subsection, we develop a motion estimation methodology that overcomes these two issues at the same time.

As stated below, \textit{\textbf{Algorithm 1}} is the main function of the proposed motion estimation methodology, where \textit{\textbf{Algorithm 2}}, \textit{\textbf{Algorithm 3}}, and \textit{\textbf{Algorithm 4}} are called in this main function.

\begin{algorithm}
\footnotesize
  
\SetAlgoLined
\KwResult{Vehicle $j$'s estimated longitudinal motion $\mathbf{\Tilde{V}_j(t)}$ and $\mathbf{\Tilde{R}_j(t)}$ in the future horizon $[t+1, t+N]$}

Vehicle $i$ associates with its target vehicle $j$, where $j=i-1$\;

\While{communication between vehicle $i$ and $j$ is currently on}{

\uIf{$n=j==0$, namely vehicle $j$ is the leader of a communication topology and does not have any target vehicle}
{Vehicle $j$ estimate its future longitudinal speed trajectory $\mathbf{\Tilde{V}_j(t)}$ based on \textbf{\textit{Algorithm 2}}\;
Vehicle $j$ cumulatively estimates its future longitudinal position trajectory $\mathbf{\Tilde{R}_j(t)}$ based on \textbf{\textit{Algorithm 3}};
}

\Else{
\For{$n==1 \rightarrow j$}{


\uIf{Vehicle $n$ is connected to its target vehicle $n-1$, namely no packet loss is in presence at this time step $t$}
{Vehicle $n$ estimates its future speed longitudinal trajectory $\mathbf{\Tilde{V}_n(t)}$ based on $\mathbf{\Tilde{V}_{n-1}(t)}$ and $\mathbf{\Tilde{R}_{n-1}(t)}$ with \textbf{\textit{Algorithm 4}}\;}
\Else{Vehicle $n$'s future speed longitudinal trajectory estimate stays the same since no information update  $\mathbf{\Tilde{V}_n(t)} = \mathbf{\Tilde{V}_n(t-1)}$\;}

Vehicle $n$ cumulatively estimates its future longitudinal position trajectory $\mathbf{\Tilde{R}_n(t)}$ based on $\mathbf{\Tilde{V}_n(t)}$ with \textbf{\textit{Algorithm 3}}\;}
Vehicle $j$'s estimated motion $\mathbf{\Tilde{V}_j(t)}$ and $\mathbf{\Tilde{R}_j(t)}$ can be derived when $n==j$;
}

Vehicle $j$ sends $\mathbf{\Tilde{V}_j(t)}$ and $\mathbf{\Tilde{R}_j(t)}$ to its following vehicle $i$;
}

\While{communication between vehicle $i$ and $j$ is currently off due to packet loss}
{
At time step $(t+k) \subseteq [t+1,t+N]$, vehicle $i$ extras $\Tilde{v}_j(t+k)$ from $\mathbf{\Tilde{V}_j(t)}$, and $\Tilde{r}_j(t+k)$ from $\mathbf{\Tilde{R}_j(t)}$, and uses them as the inputs of the motion control algorithm;
}

Vehicle $i$ disassociates with its target vehicle $j$;

\caption{\small{Main function of the motion estimation methodology, with three algorithms being called inside}}
\end{algorithm}

\textit{\textbf{Algorithm 2:}} This algorithm estimates vehicle $j$'s future longitudinal speed trajectory when it does not have any target vehicle to follow. In this case, vehicle $j$ will converge to its target speed $v_j(t \rightarrow \infty)$, which is a known and preset value. Specifically, this algorithm is proposed based upon the Intelligent Driver Model (IDM) \cite{treiber2000congested} as follows:

\begin{equation}
\Tilde{v}_j(t+k)=\Tilde{v}_j(t+k-1) + a_j^{\max } \cdot\left[1-\left(\frac{\Tilde{v}_j(t+k-1)}{v_j(t \rightarrow \infty)}\right)^{\sigma}\right] \cdot \delta t
\end{equation}

where $k \subseteq [1, N]$, and $\Tilde{v}_j(t) = v_j(t)$; $a_j^{max}$ is a preset constant denoting vehicle $j$'s maximum changing rate of longitudinal speed, which is set as 0.73 m/s$^2$ in this study; $\sigma$ is the free acceleration exponent defined by IDM, which characterizes how the acceleration of the vehicle decreases with speed, which is set as 4 in this study; $\delta t$ is the duration of a prediction time step. Based on this algorithm, the future longitudinal speed trajectory of vehicle $j$ can be given as $\mathbf{\Tilde{V}_j(t)}=\Big(\Tilde{v}_j(t+1), \Tilde{v}_j(t+2), ..., \Tilde{v}_j(t+k), ..., \Tilde{v}_j(t+N)\Big)$.

\textit{\textbf{Algorithm 3:}} This algorithm estimates vehicle $n$'s future longitudinal position trajectory $\mathbf{\Tilde{R}_n(t)}$ based on its estimated speed trajectory $\mathbf{\Tilde{V}_n(t)}$, which is given as:

\begin{equation}
\Tilde{r}_n(t+k)= \Tilde{r}_n(t+k-1) + \Tilde{v}_n(t+k-1) \cdot \delta t
\end{equation}

where $k \subseteq [1, N]$, and $\Tilde{r}_n(t) = r_n(t)$. Based on this algorithm, the future longitudinal position trajectory of vehicle $n$ can be given as $\mathbf{\Tilde{R}_n(t)}=(\Tilde{r}_n(t+1), \Tilde{r}_n(t+2), ..., \Tilde{r}_n(t+k), ..., \Tilde{r}_n(t+N))$.

\textit{\textbf{Algorithm 4:}} This algorithm estimates vehicle $n$'s future longitudinal speed trajectory when it has a target vehicle $(n-1)$ to follow, meanwhile also considers the presence of communication delay $\tau_{n(n-1)}(t+k)$. When $\tau_{n(n-1)}(t+k)<\delta t$, namely the communication delay is less than the duration of a prediction time step, then target vehicle $(n-1)$'s longitudinal speed is assumed unchanged during this delayed period:

\begin{equation}
\Tilde{v}_{n-1}(t+k)=\Tilde{v}_{n-1}\Big(t+k-\tau_{n(n-1)}(t+k)\Big)
\end{equation}

When $\tau_{n(n-1)}(t+k)>=\delta t$, namely the communication delay is equal to or longer than the duration of a prediction time step, then

\begin{equation}
\begin{split}
\Tilde{v}_{n-1}(t+k) &= \Tilde{v}_{n-1}\Big(t+k-\tau_{n(n-1)}(t+k)\Big) \\
& +\dfrac{\tau_{n(n-1)}(t+k)}{\delta t}\cdot\dot{\Tilde{v}}_{n-1}\Big(t+k-\tau_{n(n-1)}(t+k)\Big)
\end{split}
\end{equation}

In either case, target vehicle $(n-1)$'s longitudinal position can be adjusted by:

\begin{equation}
\Tilde{r}_{n-1}(t+k)=\Tilde{r}_{n-1}(t+k-1) + \Tilde{v}_{n-1}(t+k) \cdot \tau_{n(n-1)}(t+k)
\end{equation}

Then, vehicle $(n-1)$'s future longitudinal motion at each time step is used to estimate vehicle $n$'s future longitudinal speed as:

\begin{equation} \label{algorithm4}
\begin{split}
\Tilde{v}_{n}(t+k) & =\Tilde{v}_{n}(t+k-1)-\alpha_{n (n-1)} k_{n (n-1)} \\ 
& \cdot\Bigg[\bigg(\Tilde{r}_{n}(t+k)-\Tilde{r}_{n-1}\Big(t+k\Big) \\
& +l_{n-1}+\Tilde{v}_{n}(t+k) \cdot t_{n (n-1)}^{g}(t+k)\Big)\bigg) \\
& +\gamma_{n}\cdot\bigg(\Tilde{v}_{n}(t+k)-\Tilde{v}_{n-1}(t+k\Big)\bigg)\Bigg]
\end{split}
\end{equation}

Parameters of this algorithm are set according to the consensus-based motion control algorithm equation (\ref{eq:consensus}), where $i=n$ and $j=n-1$. Based on this algorithm, the future longitudinal speed trajectory of vehicle $n$ can be given as $\mathbf{\Tilde{V}_n(t)}=\Big(\Tilde{v}_n(t+1), \Tilde{v}_n(t+2), ..., \Tilde{v}_n(t+k), ..., \Tilde{v}_n(t+N)\Big)$.




\section{Case Study and Simulation Results}
\subsection{Automated Coordination at Unsignalized Intersections}
Traffic signals has been playing a crucial role in achieving safer performance at intersections. Researchers and practitioners have shown in their works that, the appropriate installation and operation of traffic signals can reduce the severity of crashes \cite{fhwa2014signalized}. However, the addition of unnecessary or inappropriately-designed signals have adverse effects on traffic safety and mobility. In addition, the dual objectives of safety and mobility conflict in many cases. The design and operation of traffic signals at intersections requires choosing elements that may lead to trade-offs in safety and mobility.

\begin{figure}[!htb]
\centering
\includegraphics[width=0.45\textwidth]{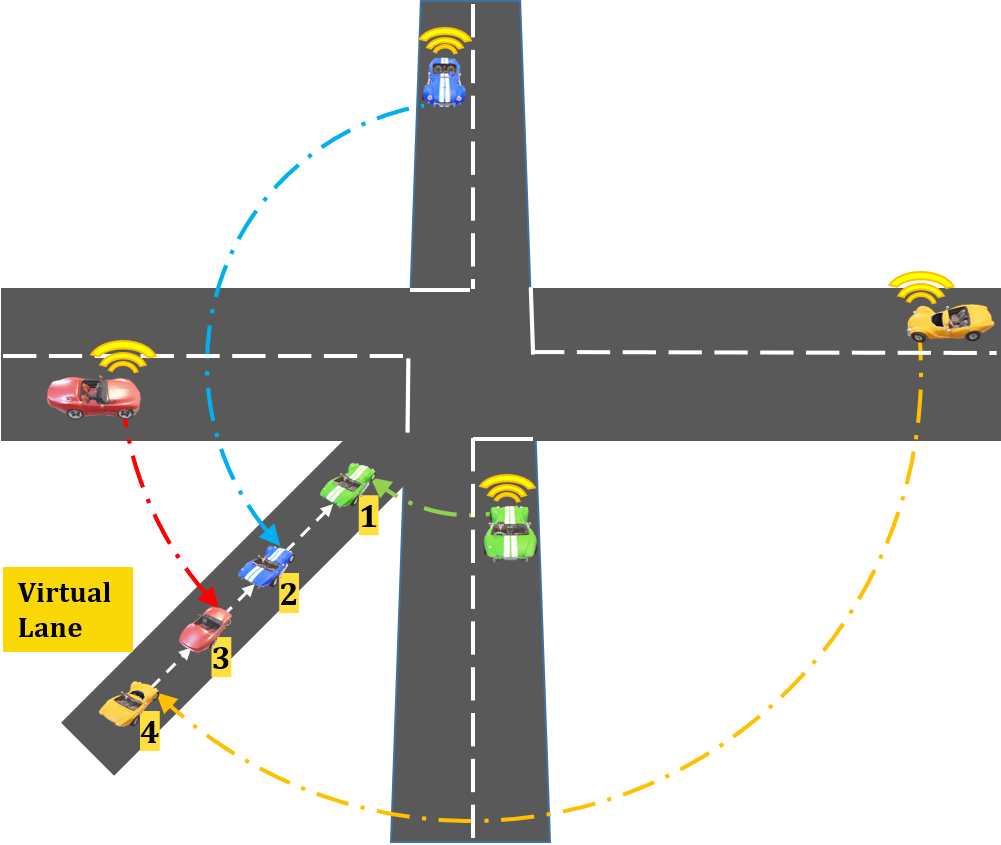}
\caption{The illustration of vehicle parameters in a vehicle-to-vehicle (V2V) communication environment.}
\label{fig:intersection}
\end{figure}

In this paper, we conduct a case study in the automated coordination scenario at unsignalized intersections, where different CAVs cooperate with each other based on V2X communications and cross the intersection without any full stop \cite{dresner2008amultiagent, guney2018scheduling}. As shown in Fig. \ref{fig:intersection}, we adopt the ``virtual vehicle'' idea that projects CAVs coming from four different legs to a virtual lane based on their longitudinal positions (more precisely, their longitudinal distances to the intersection crossing point). The consensus-based motion control algorithm (\ref{eq:consensusdelay}) can then be applied to CAVs to adjust their longitudinal positions and speeds with respect to their target vehicles (either on the same leg or another leg), and therefore form a vehicle string on the virtual lane while approaching to the intersection. In this manner, all CAVs can collaboratively cross the unsignalized intersection without any full stop.

However, V2X communications among vehicles have limited performances, with time-variant communication delay and packet loss. The proposed motion estimation methodology will estimate target CAV's motion when its information is received with delay, or not received by the ego vehicle at all.

\subsection{Game Engine Simulation Environment}
During the past decade, the rapid development of game engines makes them a popular option to model and simulate advanced vehicular technology \cite{ma2020new}. Game engines typically consist of three modules: 1) a rendering engine for 2-D or 3-D graphics, 2) a physics engine for collision detection and response, 3) and a scene graph for the management of multiple elements (e.g., models, sound, scripting, threading, etc.). They have been increasingly used by researchers to simulate autonomous driving \cite{dosovitskiy2017carla, rong2020lgsvl, wei2020end}, prototype connected vehicle systems \cite{wang2019cooperative, liu2020sensor}, study driver behaviors \cite{wang2020driver}, and etc. 

\begin{figure}[ht!]
    \centering
    \includegraphics[width=1.0\columnwidth]{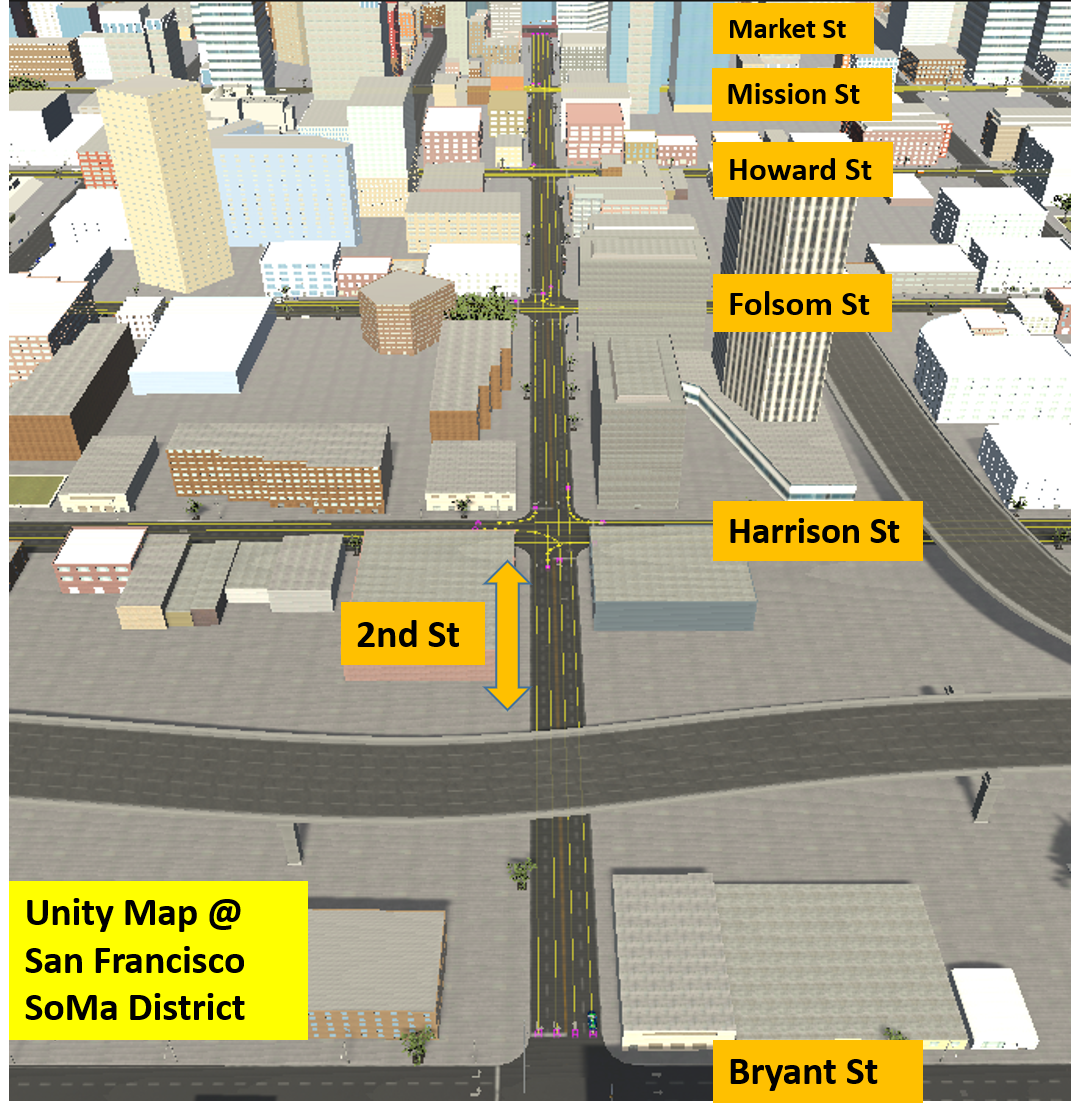}
    \caption{The map with a four-intersection (each with four legs) corridor built in Unity game engine based on the SoMa district in San Francisco}
    \label{map}
\end{figure}

In this paper, we use Unity game engine to conduct the case study of automated coordination at unsignalized intersections to evaluate our motion estimation methodology. As shown in Fig. \ref{map}, a map is built based on the South of Market (SoMa) district in San Francisco with 1:1 ratio \cite{rong2020lgsvl}. Shown as yellow lines on the road surface, centimeter-level routes along the 2nd Street, Harrison Street, Folsom Street, Howard Street, and Mission Street are further modeled in this case study, so CAVs can be located with relatively high precision. The consensus-based motion control algorithm and the proposed motion estimation methodology are applied to CAVs in this environment through Unity's C\# API.

In this simulation, we set the time-variant communication delay based on the results of our previous field test regarding 4G LTE-based V2C (not standard C-V2X) communications \cite{liao2020cooperative}. Note that this communication technology normally has higher communication delays than DSRC or C-V2X, so we adopt it in the simulation as a stress test of our motion estimation methodology. Specifically, we set the communication delay as a normal distribution with a mean value of 40 ms and a standard deviation of 0.0259. 

Additionally, we also model the packet loss in the simulation with a hybrid model, which includes both random packet loss over the whole simulation period, and certain packet loss during some specific time periods. The random packet loss modeling part is to simulate the relatively periodic communication issues, while the certain modeling part is to simulate the non-line-of-sight (NLOS) scenario when the communication is obstructed by physical objects (e.g., tall buildings or bridges). The packet loss is modelled and simulated in a more frequent manner than it is supposed to be in the real world, so we can conduct stress test of our motion estimation methodology, similar to communication delay.

\subsection{Simulation Results and Evaluation}
\begin{figure}[ht!]
    \centering
    \includegraphics[width=1.0\columnwidth]{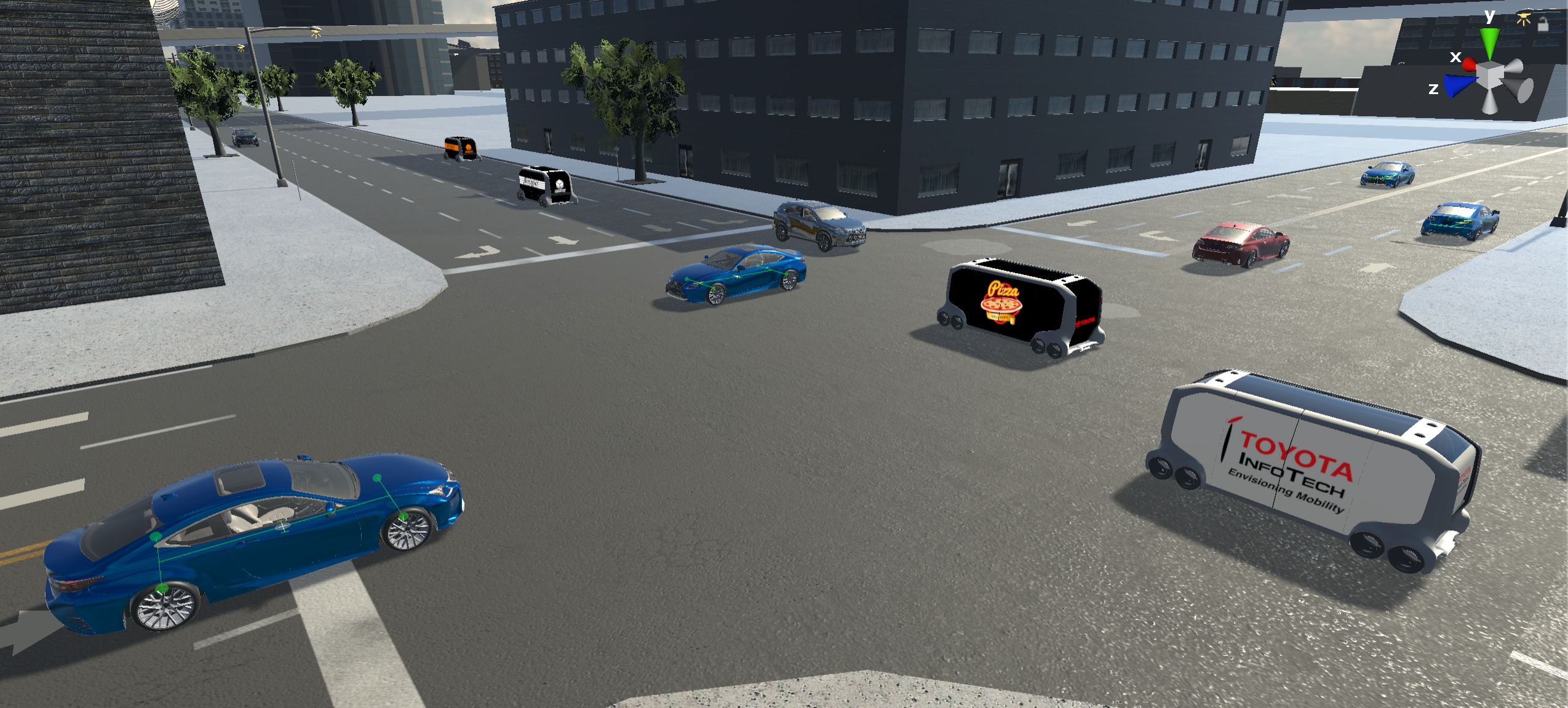}
    \caption{Unity simulation of automated coordination at an unsignalized intersection, where CAVs are implemented with the proposed motion estimation methodology to deal with communication delay and packet loss}
    \label{unity}
\end{figure}

The simulation is conducted on a Windows desktop (processor Intel Core i7-9750 @2.60 GHz, 32.0 GB memory, NVIDIA Quadro RTX 5000 Max-Q graphics card) and Unity version 2019.2.11f1. A snapshot of the simulation is shown as Fig. \ref{unity}, where CAVs are randomly generated from three legs of this 2nd-Harrison intersection (since Harrison St is a one-way street), and they automatically coordinate with each other to cross this unsignalized intersection without any collision or full stop. Besides this intersection, CAVs are also randomly generated from different legs of other three intersections along 2nd St (as illustrated in Fig. \ref{map}), so the ego vehicle on 2nd St can travel through all four unsignalized intersections in a row by automated coordination with other CAVs. 

Multiple simulation trips are conducted under various prediction time steps of the proposed motion estimation methodology, investigating the impacts on the estimation error and computational load. All CAVs in this simulation are implemented with a first-come-first-served motion planning algorithm to generate the crossing sequence, a consensus-based motion control algorithm shown earlier in this paper to generate the reference speed, and the proposed motion estimation methodology to deal with communication delay and packet loss.

\begin{figure}[!htb]
\centering
\includegraphics[width=0.45\textwidth]{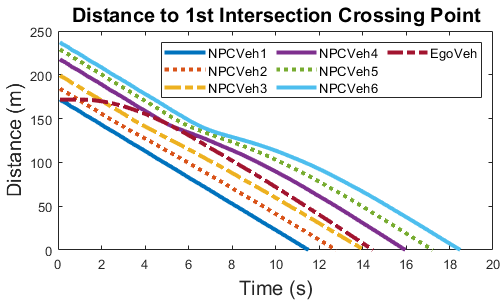}
\caption{Longitudinal position trajectory of vehicles crossing the first intersection (2nd-Harrison intersection), where the ego vehicle (dark red dashed curve) has communication delay and packet loss.}
\label{fig:distance}
\end{figure}

\begin{figure}[!htb]
\centering
\includegraphics[width=0.43\textwidth]{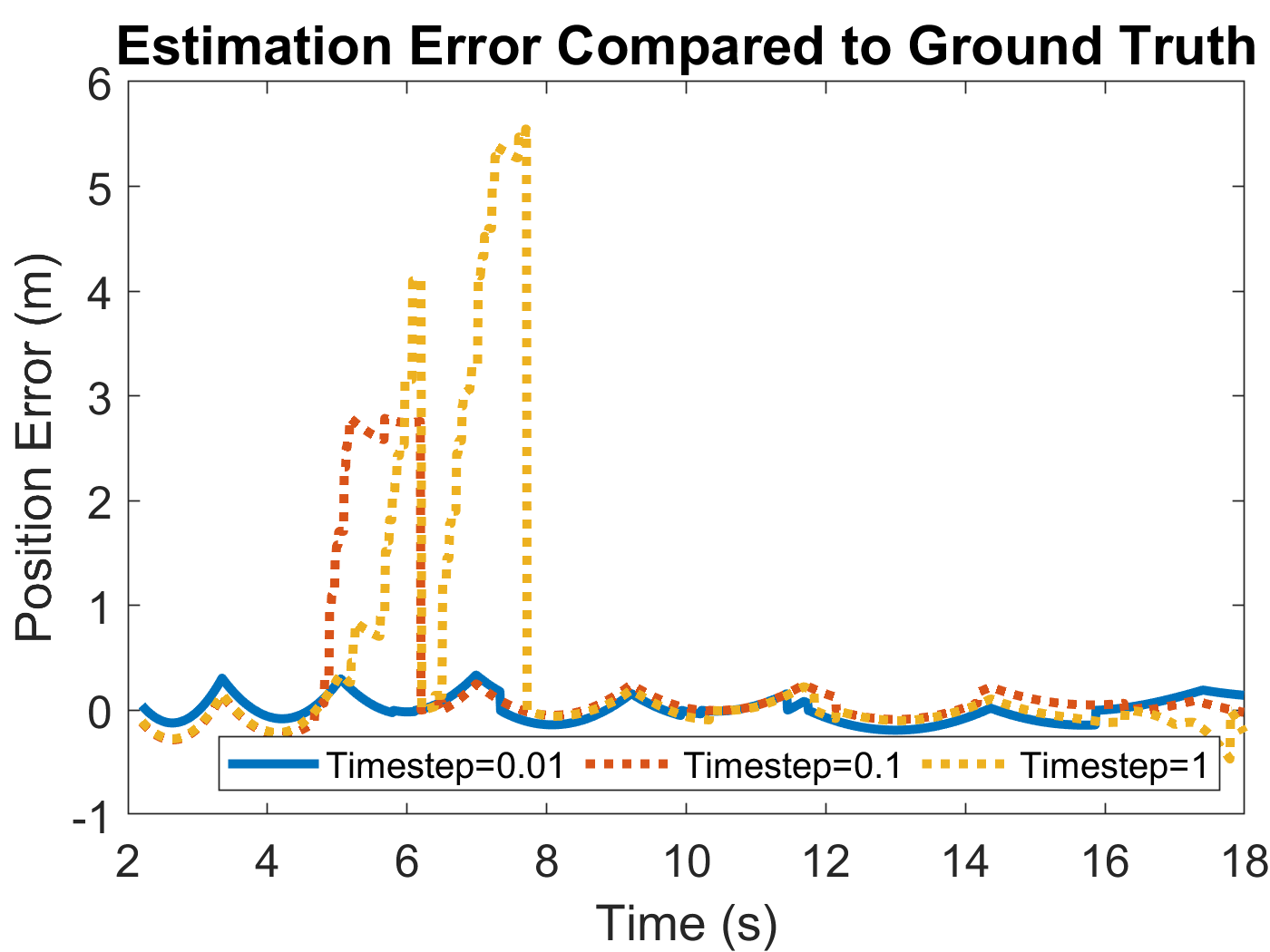}
\caption{Error of the ego vehicle's position estimation with different prediction time step values compared to the ground truth.}
\label{fig:error}
\end{figure}

\begin{figure}[!htb]
\centering
\includegraphics[width=0.3\textwidth]{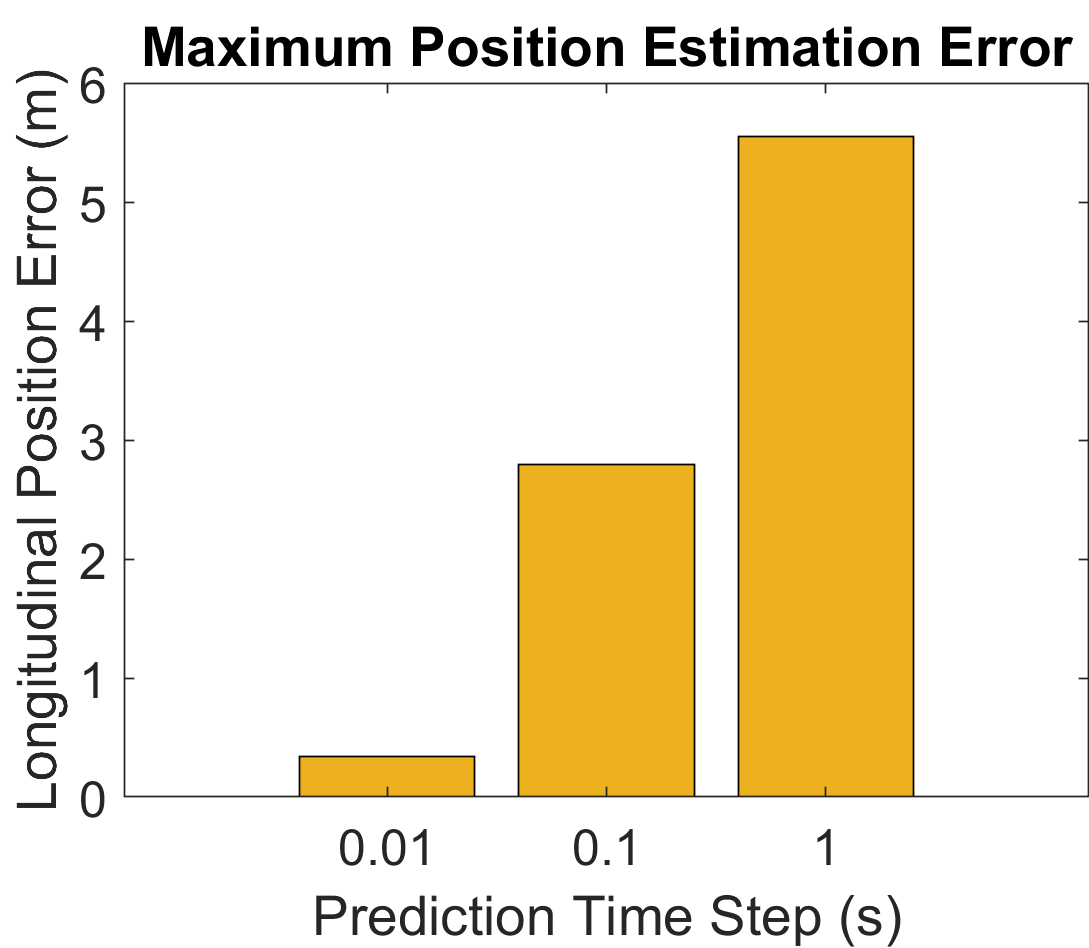}
\caption{Maximum error of the ego vehicle's position estimation with different prediction time step values.}
\label{fig:errormax}
\end{figure}

\begin{figure}[!htb]
\centering
\includegraphics[width=0.3\textwidth]{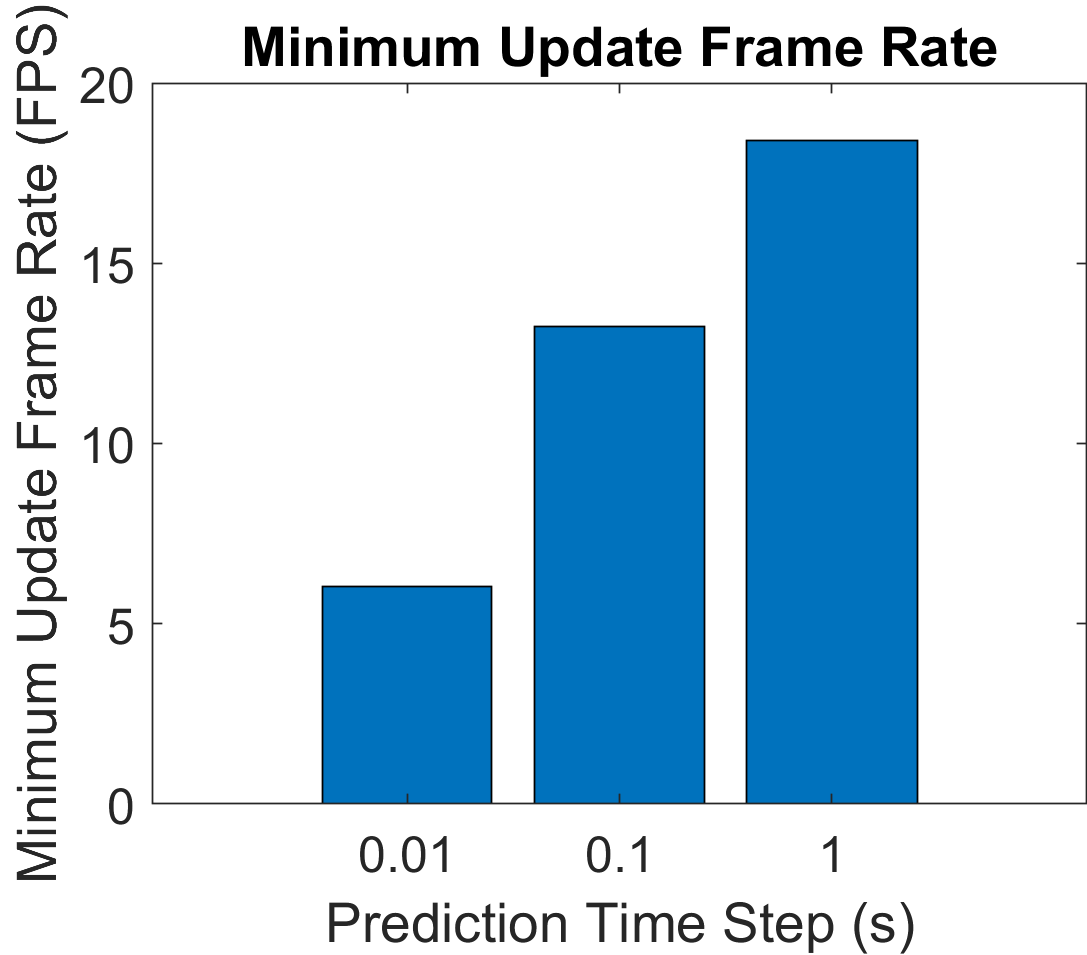}
\caption{Minimum update frequency with different prediction time step values.}
\label{fig:FPS}
\end{figure}

As can be seen from the results, Fig. \ref{fig:distance} shows the ground-truth longitudinal position trajectory of CAVs crossing 2nd-Harrison intersection, which is the first of four consecutive intersections in the simulation network. The ego vehicle is applied with the time-variant communication delay and the hybrid packet loss introduced in the previous subsection. Specifically, two major periods of packet loss happen during 4-6 second and 6-8 second, respectively, where NPC vehicle 4 (which considers the ego vehicle as its target vehicle) gets a little bit close to the ego vehicle due to the estimation error (shown in Fig. \ref{fig:error}). However, this does not cause any rear-end collisions in the simulation, since these two vehicles do not travel on the same leg of this intersection. 

Besides aforementioned two major periods of packet loss during 4-8 second, estimation error can also be observed during the whole period when the ego vehicle crosses the first intersection, as shown in Fig. \ref{fig:error}. These errors are generated by the combination of periodic packet loss and time-variant communication delay, and are shown to be capped by $\pm 0.5$ m with the help of the proposed motion estimation methodology.

Additionally, the impacts of different values of the prediction time step $\delta t$ are also investigated. Shown in Fig. \ref{fig:error} and Fig. \ref{fig:errormax}, when the prediction happens more frequently, the maximum position estimation error is decreased. Particularly, when the prediction time step is 0.01 s (i.e., prediction frequency is 100 Hz), the estimation error is always less than 0.2 m based on the simulation setting of communication delay and packet loss. However, when the prediction time step is 1 s (i.e., prediction frequency is 1 Hz), the estimation error can reach 5.8 m during the major packet loss. This is caused by the difference of information granularity, where a more frequent prediction also means a more frequent information update, which could happen immediately after the packet resumes to be received.

However, a more frequent prediction also leads to a higher computational load. Since this simulation is conducted in Unity game engine, the computational load is measured by the minimum update frame rate of the simulation. As shown in Fig. \ref{fig:FPS}, When the prediction time step is 0.01 s, the frame rate drops to a minimum of 6 FPS, indicating the simulation can be run only six frames per second at that time instant (with the simulation hardware setting). However, when the prediction time step is 1 s, the frame rate is always higher than 19 FPS. This is caused by the difference of iterations that the proposed motion estimation methodology needs to be called. However, since this simulation is conducted on a single computer for multiple vehicles as a centralized fashion, the concern of computational load can be further relieved if distributed computing can be adopted by each individual vehicle in the future.

\section{Conclusion}
\label{sec:concl}
In this paper, we have investigated the issues of communication delay and packet loss in V2X communications among CAVs. A motion estimation methodology has been developed based on the consensus-based feedforward/feedback motion control algorithm, which handles aforementioned two communication issues at the same time. A case study of automated coordination at unsignalized intersection has been conducted, where CAVs coordinate with each other through V2X communications to cross the intersection without any full stop. The simulation study in Unity game engine has shown that, the proposed motion estimation methodology (with 0.01 s prediction time step) can cap the position estimation error at $\pm 0.5$ m with the presence of periodic packet loss and time-variant communication delay. Additionally, different impacts of the prediction time steps on the estimation error and computational load have also been studied in the simulation.

One major future step of this work is to implement the proposed motion estimation methodology on real CAVs, and conduct real-world field implementations. Real vehicle dynamics, instead of the simplified version adopted in this paper, needs to be considered to improve the fidelity of the algorithm. Additionally, data measured by CAVs' on-board perception sensors can also be leveraged to provide additionally sources of motion estimation, which could significantly decreases the motion estimation error of the proposed methodology.

\bibliographystyle{ieeetr} 
\bibliography{sampleBib}

\section{Contact Information}
Ziran Wang, Ph.D. \newline
Research Scientist \newline
Toyota Motor North America R\&D, InfoTech Labs \newline
Work Address: 465 N Bernardo Ave, Mountain View, CA 94043 \newline
Work Phone: (650) 439-9524 \newline
Work Email: ziran.wang@toyota.com

\section{Acknowledgments}
This work is conducted under the ``Digital Twin'' project at Toyota Motor North America R\&D. The authors would like to thank the feedback and guidance from Dr. Nejib Ammar and Dr. Bin Cheng to improve this work. 

The contents of this paper only reflect the views of the authors, who are responsible for the facts and the accuracy of the data presented herein. The contents do not necessarily reflect the official views of Toyota Motor North America R\&D, InfoTech Labs.

\section{Definitions, Acronyms, Abbreviations}

\begin{table}[ht]
\centering
\begin{tabular}{L{0.1\textwidth} L{0.33\textwidth}}
\textbf{CAV} & Connected and Automated Vehicle\\
\textbf{DSRC} & Dedicated Short Range Communications \\
\textbf{C-V2X} & Cellular Vehicle-to-Everything \\
\textbf{V2V} & Vehicle-to-Vehicle \\
\textbf{V2I} & Vehicle-to-Infrastructure \\
\textbf{V2N} & Vehicle-to-Network \\
\textbf{V2C} & Vehicle-to-Cloud \\
\textbf{V2P} & Vehicle-to-Pedestrian \\
\textbf{ACC} & Adaptive Cruise Control \\
\textbf{CACC} & Cooperative Adaptive Cruise Control \\
\textbf{NLOS} & Non-Line-of-Sight \\
\textbf{FPS} & Frames Per Second
\end{tabular}
\end{table}

\end{document}